\documentclass[11pt]{article}
  
\usepackage{amsmath,amssymb}

\usepackage{epsfig}  
\usepackage{graphicx}               
\usepackage{url}

\newcommand{\nc}{\newcommand}  

\nc{\beq}{\begin{equation}}  
\nc{\eeq}{\end{equation}}  
\nc{\beqa}{\begin{eqnarray}}  
\nc{\eeqa}{\end{eqnarray}}  
\nc{\bea}{\begin{eqnarray}}  
\nc{\eea}{\end{eqnarray}}  
\nc{\ra}{\rightarrow}  
\nc{\lsim}{\begin{array}{c}\,\sim\vspace{-21pt}\\< \end{array}}  
\nc{\gsim}{\begin{array}{c}\sim\vspace{-21pt}\\> \end{array}}  
\nc{\Tr}{{\rm Tr}}
\nc{\slsh}{\slash\hspace*{-0.22cm}}

\def\ie{{\it i.e.}}
\def\be{\begin{equation}}
\def\ee{\end{equation}}
\def\bea{\begin{eqnarray}}
\def\eea{\end{eqnarray}}

\newcommand{\bra}[1]{ \langle {#1} | }
\newcommand{\ket}[1]{ | {#1} \rangle }

\newcommand{\kev}{{\rm keV}}

\newcommand{\gev}{{\rm GeV}}

\newcommand{\mev}{{\rm MeV}}

\title{  
\vspace*{-2.3cm}  
\begin{flushright}  
\normalsize{  
FERMILAB-PUB-10-143-T 
  }  
\end{flushright}  
\vspace{1.5cm}  
\Large  
\textbf{The Tevatron at the Frontier of Dark Matter Direct Detection
}\vspace*{1.0cm}   
}
\author{\large
Yang Bai, Patrick J. Fox and Roni Harnik
\\ \vspace{0.5cm} \\
\normalsize\emph{Fermi National Accelerator Laboratory,  
P.O. Box 500, Batavia, IL 60510, USA} 
}
\date{}

\begin{document}  
\setcounter{page}{0}  
\maketitle  

\vspace*{1cm}  
\begin{abstract} 
Direct detection of dark matter (DM) requires an interaction of dark matter particles with nucleons. The same interaction can lead to dark matter pair production at a hadron collider, and with the addition of initial state radiation this may lead to mono-jet signals. Mono-jet searches at the Tevatron can thus place limits on DM direct detection rates.  We study these bounds both in the case where there is a contact interaction between DM and the standard model and where there is a mediator kinematically accessible at the Tevatron.  We find that in many cases the Tevatron provides the current best limit, particularly for light dark matter, below $\sim 5$ GeV, a and for spin dependent interactions. Non-standard dark matter candidates are also constrained. The introduction of a light mediator significantly weakens the collider bound. A direct detection discovery that is in apparent conflict with mono-jet limits will thus point to a new light state coupling the standard model to the dark sector. Mono-jet searches with more luminosity and including the spectrum shape in the analysis can improve the constraints on DM-nucleon scattering cross section.  
\end{abstract}  
\thispagestyle{empty}  
\newpage  
  
\setcounter{page}{1}

\baselineskip18pt   

\section{Introduction}
\label{sec:intro}

From astronomical and cosmological observations it is now clear that $\sim 25\%$ of the matter-energy content of the universe if made up by dark matter (DM).  Although DM has so far only been observed through its gravitational interactions the quest for a more direct observation of DM is taking place simultaneously on many fronts.  Indirect searches look for signals of standard model (SM) particle production from DM annihilations in our galaxy, direct searches look for interactions of DM with SM particles in underground detectors and colliders attempt to produce the DM and measure it.  We will concentrate here on direct detection and collider searches.

If dark matter is to be observed in direct detection searches it must couple to quarks or gluons~\footnote{DAMA is an exception as, unlike other experiments, it does not distinguish between nuclear and electron recoils of DM.}. The same couplings lead to direct DM production at hadronic colliders such as the Tevatron, and we wish to investigate the connection between the two types of search.  We will do so in a model independent fashion~\cite{Birkedal:2004xn}; we will assume that the DM is fermionic and that there is some massive state whose exchange couples DM to quarks. The mediator may be a SM gauge boson, the Higgs or a new particle (if the new particle is very heavy we can describe its effects with an effective contact operator).
Although the processes that give direct detection and those that give DM production occur through s- and t-channel exchange  of the same mediator, the regimes probed in the two types of experiment are very different.  The momentum exchange during a DM-nucleus recoil is $\sim 100\ \mev$ whereas at the Tevatron the typical momentum exchange is $10-100\ \gev$.  This leads to two interesting regimes to consider when comparing bounds from the two types of experiments: heavy mediators $M\gtrsim  100\ \gev$ and light mediators $M\lesssim 100\ \gev$.

The momentum exchange at direct detection experiments is sufficiently low that for all but the lightest mediators below ${\cal O}(100~\mbox{MeV})$, which we do not consider here, the mediator can effectively be integrated out and the scattering rate in both regimes scales as,
\be
\sigma_{\rm DD}\sim g_\chi^2\, g_q^2\, \frac{\mu^2}{M^4}\,,
\ee
where, for simplicity, we have ignored form factors and possible momentum and velocity dependence in the cross section. Here, $g_\chi$ and $g_q$ are couplings of the mediator to DM and quarks. $\mu$ is the reduced mass of the DM-nucleon system. 

In contrast the two regimes behave very differently at colliders.  Concentrating on direct production of a pair of DM particles and an initial state emission of a jet, we estimate the mono-jet + $\slsh{E_T}$ partonic production cross section in the two cases to be
\be
\sigma_{1j} \sim 
\begin{cases}
\alpha_s \,g_\chi^2\, g_q^2\,\frac{1}{p_T^2} & \quad M\lesssim p_T \,, \vspace{3mm}\\
\alpha_s \,g_\chi^2\, g_q^2\, \frac{p_T^2}{M^4} &\quad  M\gtrsim p_T \,, \\
\end{cases}
\label{eq:colliderqual}
\ee
where $\alpha_s$ is the QCD coupling and $p_T$ is the transverse momentum of the jet, which is typically $\sim\mathcal{O}(100)\,\gev$ at the Tevatron\footnote{Note that (\ref{eq:colliderqual}) is only qualitative in nature.  The limits are correct for mediator masses well above and below the $p_T$ of the jet.}.  Thus, for the heavy mediator case the (partonic) production cross section at the Tevatron, where $p_T\sim 100\ \gev$, is $\mathcal{O}(1000)$ times larger than the direct detection cross section for $\mu \sim 1$~GeV when the DM is heavier than the nucleon mass.  The CDF mono-jet search~\cite{CDFmonojet} analysed $\sim1\ {\rm fb}^{-1}$ and saw no significant discrepancy from the SM, thus limiting the DM + mono-jet production cross section to be smaller than $\sim 500\ {\rm fb}$.  Due to the factor of 1000 mentioned above, this will translate to bounds in the neighborhood of $0.5$ fb in direct detection experiments, the exact bound at direct detection experiments will depend upon the details of the parton density functions relating the partonic cross section of (\ref{eq:colliderqual}) to the actual CDF mono-jet bound.

This is to be compared with direct detection current searches.  
Null results from experiments such as CDMS~\cite{Ahmed:2009zw}, XENON\cite{Angle:2007uj,Aprile:2010um} and others,
 place strong constraints on the cross section of DM to recoil from a nucleus, $\sigma \lesssim 10^{-3}-10^{-4}~{\rm fb}$ for a 10-100 GeV WIMP scattering elastically through a spin independent (SI) interaction.  Thus, for this situation it seems that direct detection has greater reach.  However, due to the threshold to detect a DM recoil in these experiments there is a DM mass below which these experiments are no longer sensitive, typically this lower bound is $m_\chi\sim 5-10\ \gev$, there is no such threshold in collider searches.  

Furthermore, the DAMA collaboration~\cite{Bernabei:2008yi} have observed a signal consistent with DM scattering from NaI 
which is inconsistent with bounds on a standard WIMP from CDMS and other experiments.  
This has motivated the introduction of non-standard DM scenarios that can make these seemingly discrepant results consistent.  The cross sections necessary to explain DAMA are considerably larger than $10^{-3} {\rm fb}$ and may allow these scenarios to be probed directly at the Tevatron, due to the increase in cross section described above.  
Another possibility that has been motivated both by DAMA and the recent CoGeNT~\cite{Aalseth:2010vx} excess is that dark matter is light, below about 10 GeV, and is thus transfers small momenta to nuclei giving a signal near threshold. The Tevatron will place a strong bound for dark matter particles below 5 GeV.
Finally, spin-dependent (SD) WIMP-nucleus scatterings are not coherent and therefore are not enhanced by an $A^2$ factor.  Typical bounds on a SD WIMP-proton scatter from direct detection are $\sim 1$ fb , and will be severely impacted by the mono-jet bounds presented here.

We will begin our discussion with a model independent operator analysis, corresponding to very heavy mediation particles (such as a heavy $Z'$ or squarks). In Section~\ref{sec:Ops} we will introduce some representative four fermion operators supressed by a cutoff scale. We will then place limits on the strengths of these operators from the Tevatron mono-jet search. In Section~\ref{sec-DD} we will translate the Tevatron bounds to limits on direct detection cross section for different dark matter scenarios. In Section~\ref{sec:lightmediator} we move on to introduce lighter mediators that are kinematically accesible at the Tevatron and find that these can either slightly enhance or severely weaken the Tevatron bounds. In Section~\ref{sec-discussion} we will discuss possible enhancements to the Tevatron dark matter search using the mono-jet $p_T$ spectrum, and conclude.

\section{Operators and mono-jets}
\label{sec:Ops}

Throughout this paper, we will assume a dark matter particle, $\chi$, as a Dirac fermion. The operators we will study are,
\bea\label{eq:Ops}
\mathcal{O}_1 &=& \frac{i\,g_\chi\,g_q}{q^2-M^2}\left(\bar{q}q\right) \left( \bar{\chi}\chi\right)  \,,\nonumber \\
\mathcal{O}_2 &=& \frac{i\,g_\chi\,g_q}{q^2-M^2}\left(\bar{q}\gamma_\mu q\right) \left( \bar{\chi}\gamma^\mu\chi\right) \,, \nonumber \\
\mathcal{O}_3 &=& \frac{i\,g_\chi\,g_q}{q^2-M^2}\left(\bar{q}\gamma_\mu\gamma_5 q\right) \left( \bar{\chi}\gamma^\mu\gamma_5 \chi\right)\,,   \nonumber \\
\mathcal{O}_4 &=& \frac{i\,g_\chi\,g_q}{q^2-M^2}\left(\bar{q}\gamma_5 q\right) \left( \bar{\chi} \gamma_5\chi\right) \,,
\eea
Here we take $q=u,d,s$ and turn on each operator one at a time (but results for a flavor universal operator will be easy to deduce). 
 $q^2$ is the exchanged momentum and the suppression scale $M$ is related to the mass of the particle whose exchange generates the four fermion operator.  
 
This is a representative set of operators that will generate a variety of dark matter scattering scenarios. 
Majorana dark matter will yield similar result (though for a Majorana spinor there are no vector interactions). 
 Initially we will assume that the mediator is heavy and integrate it out, but in Section~\ref{sec:lightmediator} we will discuss the effect of a light mediator. There are two additional operators $\bar{\chi}\sigma^{\mu\nu}\chi\,F_{\mu\nu}$ and $H^\dagger H\bar{\chi}\chi$ appearing up to the dimension six level. While they are less constrained at the Tevatron, we leave their study and the study of operators involving the three heavy quark flavors to future work\footnote{There is also considerable uncertainty in the heavy quark content of the nucleons for several of the operators under discussion~\cite{Song:2001ri}.}. 
 
Operator $\mathcal{O}_1$ leads to spin-independent coupling between the DM and a nucleus and can be thought of as arising from exchange of a scalar of mass $M$, $\mathcal{O}_2$ is similar but occurs through vector exchange.  Operator $\mathcal{O}_3$ is generated through axial-vector exchange and gives a spin-dependent coupling, and $\mathcal{O}_4$ could arise from exchange of a pseudo-scalar and gives a momentum dependent and spin-dependent DM coupling. Various combinations of these operators may be also generated by mediators charged under the SM such as squarks in supersymmetry.
\begin{figure}[t]
\begin{center}
\includegraphics[width=0.48\textwidth]{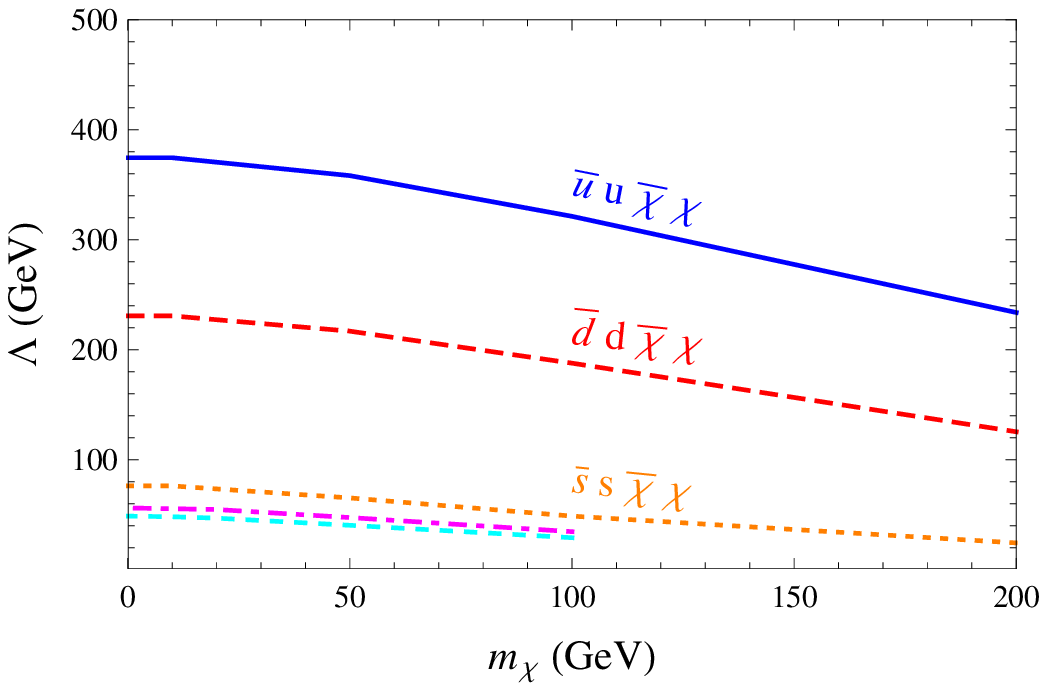} \hspace{2mm}
\includegraphics[width=0.48\textwidth]{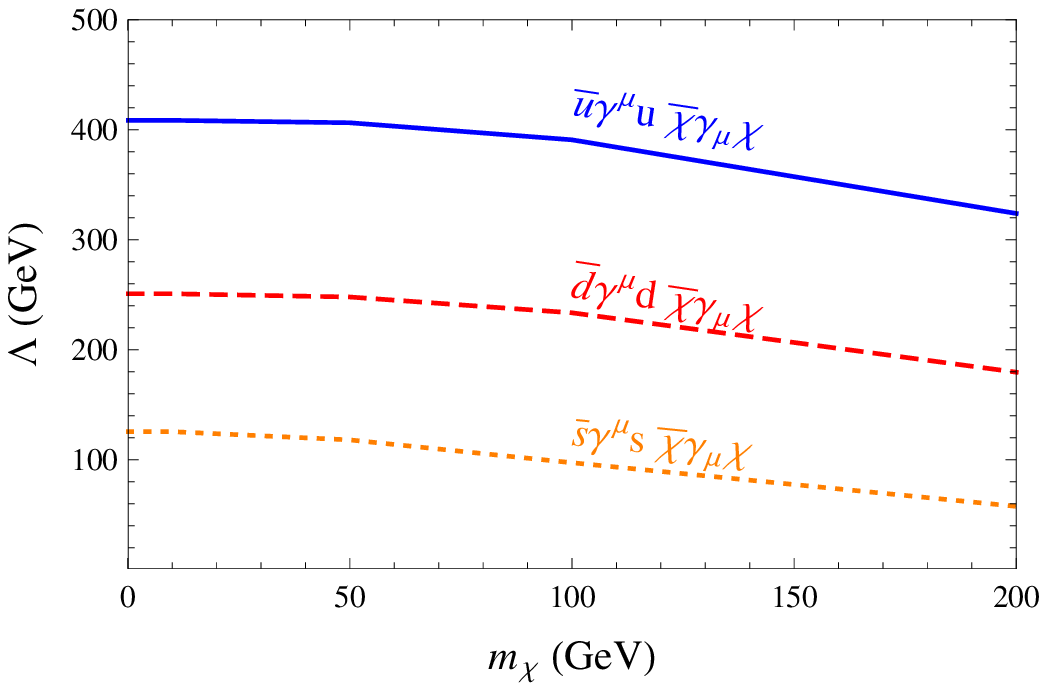} \\ \vspace{2mm}
\includegraphics[width=0.48\textwidth]{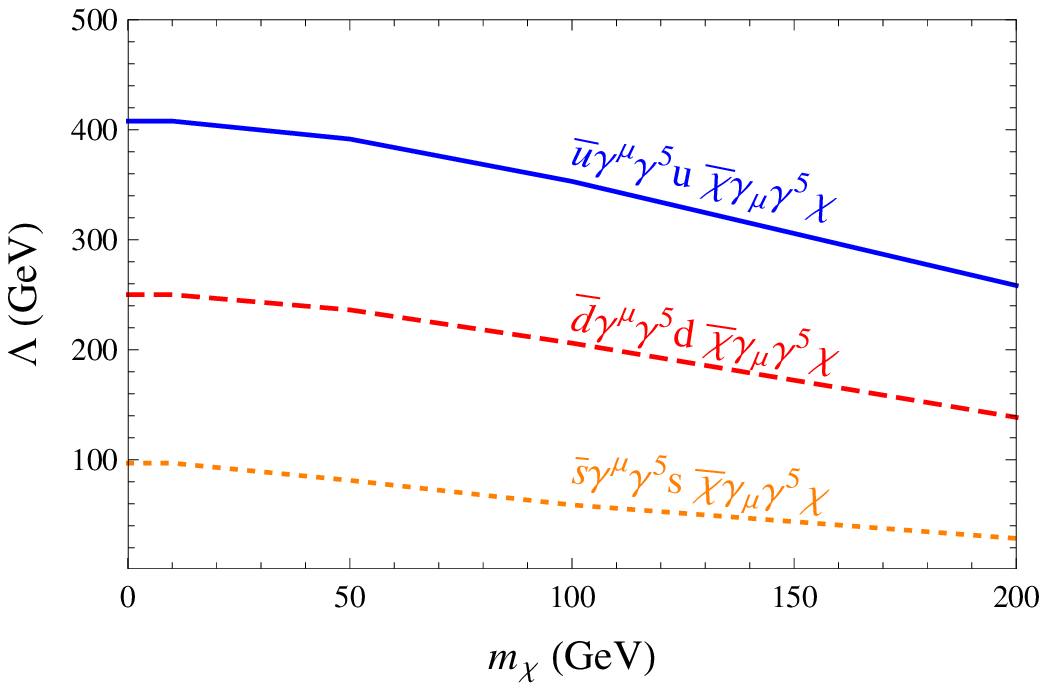} \hspace{2mm}
\includegraphics[width=0.48\textwidth]{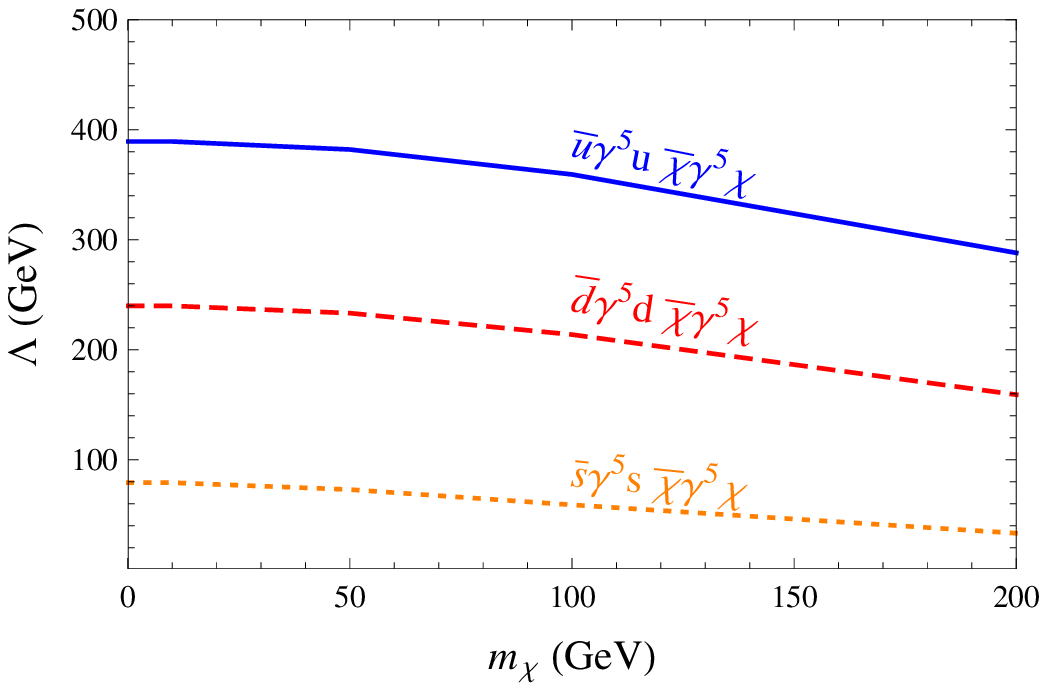}\\ \hspace{2mm}
\caption{The constraints on the cutoffs of different operators from the CDF mono-jet search data at 90\% C.L..}
\label{fig-cutoff}
\end{center}
\end{figure}
\subsection{Tevatron limits}

The CDF collaboration has performed a search for one jet events with large missing transverse energy using 1 fb$^{-1}$ of data~\cite{CDFmonojet}.  CDF considered events with
a leading jet $p_T$ and missing transverse energy both greater than 80 GeV. Events with a second jet with a $p_T< 30$ GeV were included but events with additional jets with transverse energy above 20 GeV were not. The number of observed events was 8449, a slight deficit compared to an expected background of 8663$\pm$332.
The standard model backgrounds are dominated by $Z$+jet, $W$+jet with a missed lepton. QCD and ``non-collision'' background events contribute subdominantly to the background, but due to their high uncertainty they add a significant portion to the uncertainty of the background. 
The $p_T$ spectrum observed by CDF compares well with the expected background, however since the background uncertainty was only presented for the total number of events we will only use a simple counting experiment to place bounds. We will elaborate more on the potential of a multi-bin analysis in Section~\ref{sec-discussion}.

We have generated signal events for the operators ${\cal O}_{1-4}$ using calcHEP~\cite{Pukhov:2004ca}  introducing a single operator at a time. We work with leading order parton level events, but have checked that including showering, hadronization and detector effects does not change our results significantly, see Section~\ref{sec-discussion}. Only the 80 GeV $p_T$ cut on the leading jet need be applied since we only generate leading order events and no additional jets are present. 
We add the signal events to the standard model expectation and require that the total number of signal plus background events do not disagree with the observation at more than the 90\% confidence level.
This allows us to place limits on the cutoff $\Lambda \equiv M/\sqrt{g_\chi g_q}$  which are shown in Figure~\ref{fig-cutoff}. 

Note that the bounds on the various types of operators are remarkably similar, yet at low energies they will produce radically different direct detection signals. 
As expected, the Tevatron bounds are rather flat in energy for low dark matter mass, and start becoming weaker above 50 GeV, where kinematics and PDF suppression starts entering. Due to PDF's of different quarks, the up-quark operators always provide the most stringent bound. The constraints on the strange-quark operators are fairly weak, which justifies neglecting of even heavier quark operators. 
We note that, when the cutoff is close to the dark matter mass, the real constraints are subject to order one changes and we shall investigate this region further in Section~\ref{sec:lightmediator}.

\section{Direct detection limits} \label{sec-DD}

In this section we translate the bound on the strength of the various operators into bounds on direct detection scattering rates. This enables us to plot Tevatron limits in the standard $\sigma - m_\chi$ plane. As we shall see, the Tevatron places competitive direct detection bounds for several interesting scenarios. We will summarize some of the interesting results here before presenting them in greater detail.
In general collider bounds will be relatively more powerful in cases where direct detection scattering is suppressed, either by kinematics such as when dark matter is very light such as~\cite{Petriello:2008jj,Bottino:2009km,Kuflik:2010ah,Feldman:2010ke,Graham:2010ca,Essig:2010ye} and the case of inelastic scattering~\cite{TuckerSmith:2001hy}, or by other suppressions such as when dark matter scattering is spin dependent or momentum dependent~\cite{Feldstein:2009tr,Chang:2009yt}.  This is because none of these suppressions will be effective for collisions at high energies.  

For instance the $\sim 100\ \kev$ splitting between incoming and outgoing DM states in inelastic dark matter is not an impediment to their production at colliders.  In addition, the Tevatron is not limited by the features of the DM velocity distribution in our galaxy and so is able to probe down to very low dark matter masses, well below the thresholds of direct detection experiments. 
Colliders will also be powerful in constraining direct detection scenarios in which a small fraction of dark matter is participating in scattering, but with an enhanced rate (such as~\cite{Essig:2010ye} though in this particular model the interaction is mediated by a light boson which relaxes the bound).  Additionally, vector couplings that are purely to strange or other sea quarks are not accessible to direct detection experiments and collider bound may be the only way to discover them experimentally.

In our analysis we place explicit limits on single flavor operators, we comment on the bounds on combinations of operators at the end of this section.  Flavor universal operators will have results that are close to the best single flavor operator (typically that of up).  We note in passing that mono-jet searches will not be able to constrain models where the DM-SM coupling does not involve two neutral states from the dark sector.  For instance resonant dark matter (rDM) \cite{Bai:2009cd} involves the DM state and a nearby charged state and so its Tevatron signal would instead be a jet, missing energy and a charged track.

\subsection{Spin independent}

The operators $\mathcal{O}_1$ and $\mathcal{O}_2$ induce a spin independent scattering of dark matter off of nuclei. To compute this scattering cross section off a nucleon, $N=p,n$, we will need to know the quark content of the nucleon $\bra N  \bar{q}\,\Gamma\, q \ket N$ for each of these operators.  At the nucleon level these operators become
\bea
\mathcal{O}^{Nq}_{1} &=&  B^N_q \frac{\left(\bar{N}N\right) \left( \bar{\chi}\chi\right)}{\Lambda^2} \,, \\
\mathcal{O}^{Nq}_2 &=& f^N_q \frac{\left(\bar{N}\gamma^\mu N\right) \left( \bar{\chi}\gamma_\mu\chi\right)}{\Lambda^2} ~,\nonumber 
\eea
we consider the case $M^2\gg q^2$ and $\Lambda=M/\sqrt{g_\chi g_q}$.
The coefficients necessary to translate the quark level operators to the nucleon operators are given by \cite{Cheng:1988im,Belanger:2008sj,Barger:2008qd}
\bea
B_u^p & = B_d^n & = 8.22 \pm 2.26\,, \nonumber \\
B_d^p & = B_u^n & = 6.62 \pm 1.92\,, \nonumber \\
B_s^p & = B_s^n & = 3.36 \pm 1.45
\eea
In extracting these conversion factors we have used the quark masses ratios $m_u/m_d=0.553\pm 0.043$, $m_s/m_d=18.9\pm0.8$ and quark mass $m_s=105\pm 25\ \mev$
\cite{Amsler:2008zzb}.

\begin{figure}[t]
\begin{center}
\includegraphics[width=0.48\textwidth]{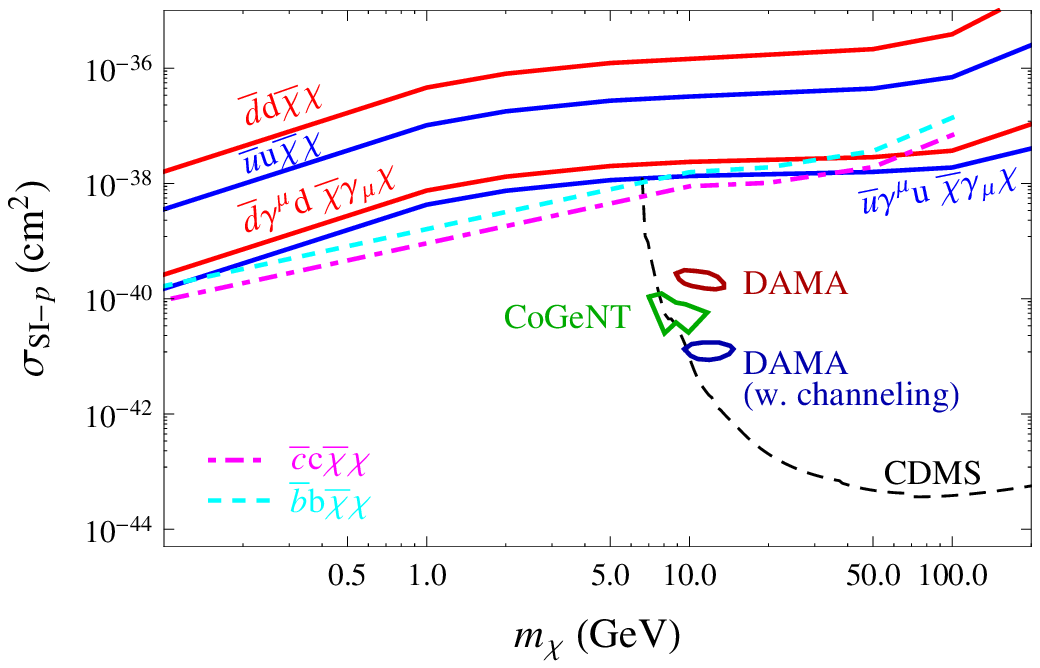} \hspace{2mm}
\includegraphics[width=0.48\textwidth]{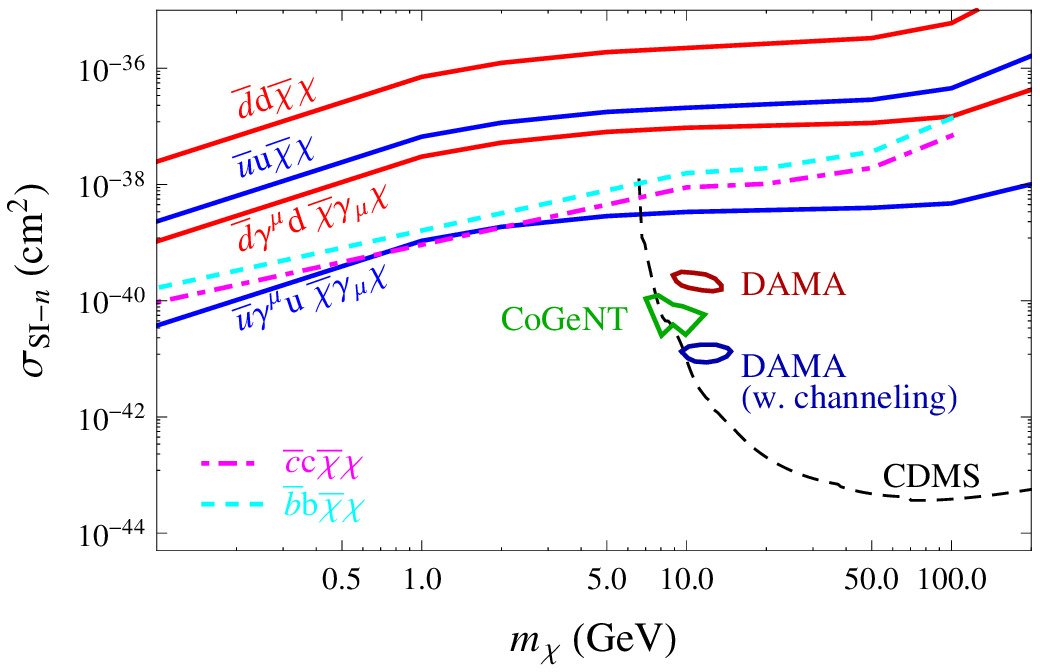}
\caption{Left panel: the constraints on the spin-indepedent DM-proton scattering cross section. Relevant experimental bounds are shown as labeled. Right panel: the same as the left panel but for the constraints on the spin-indepedent DM-neutron scattering cross section.}
\label{fig-SI-limits}
\end{center}
\end{figure}

For the vector operator, $\mathcal{O}_2$, $f_u^p=f_d^n=2$ and $f_d^p=f_u^n=1$ and for all other quarks $f=0$.   Note this means that if the DM couples through vector couplings to second and third generation quarks only then it can \emph{never} be discovered in direct detection experiments, but can be found using colliders.  At low DM speed the leading contributions to the scattering cross section in each case are 
\bea
\sigma^{Nq}_{1}&=& \frac{\mu^2}{\pi\Lambda^4}\, B_{Nq}^2  \,,\\
\sigma^{Nq}_{2}&=& \frac{\mu^2}{\pi\Lambda^4}\, f^2_{Nq} \,,
\eea
where $\mu$ is the reduced mass of the dark matter-nucleon system. The Tevatron limits on spin independent dark matter scattering for the various operators is shown in Figure~\ref{fig-SI-limits}.
The recent results from CoGeNT~\cite{Aalseth:2010vx}, CDMS~\cite{Ahmed:2009zw} and DAMA~\cite{Bernabei:2008yi} with and without channeling are also shown in Figure~\ref{fig-SI-limits}. Note that the limits are slightly different for protons and neutrons simply because they are derived from proton rather than neutron collisions. The up-type and vector coupling operator are the most constrained operators. 
For dark matter with a mass below around 5 GeV, the mono-jet searches at CDF provide the world-best spin-independent bound.

\subsection{Spin dependent}

Models in which dark matter scattering is spin dependent are even more constrained by collider experiments. This is because SD scattering is suppressed relative to SI at low momentum transfer, because the scattering is not coherent over the whole nucleus, while there is no relative suppresion between the two at high energies. Of the operators under consideration, spin dependent scattering is caused by the axial vector operator $\mathcal{O}_3$. For a complete list of all operators, see~\cite{Agrawal:2010fh}.

\begin{figure}[t]
\begin{center}
\includegraphics[width=0.48\textwidth]{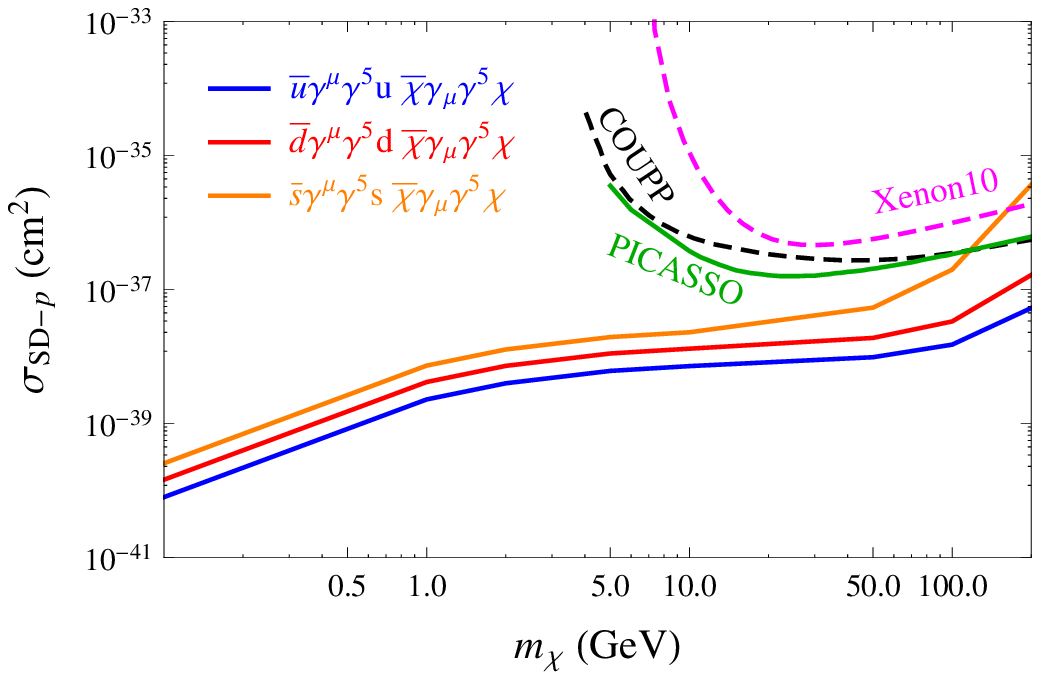}\hspace{2mm}
\includegraphics[width=0.48\textwidth]{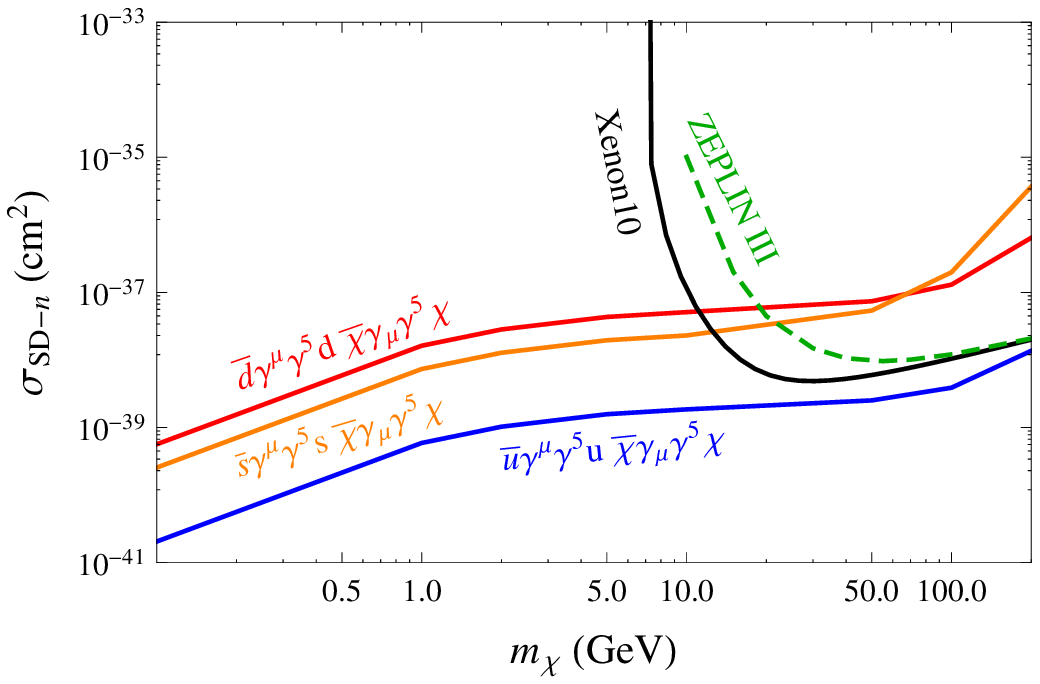}
\caption{Left panel: the constraints on the spin-dependent DM-proton scattering cross section for the up, down and strange (bottom to top solid lines) axial-vector operators.  Relevant experimental bounds are also shown. Right panel: the same as the left panel but for the constraints on the spin-indepedent DM-neutron scattering cross section.}
\label{fig-SD-limits}
\end{center}
\end{figure}

Again, in order to compute the DM scattering cross section off a nucleon, $N=p,n$, we will need $\bra N  \mathcal{O}_3 \ket N$, leading to
\beq
\mathcal{O}^{Nq}_3 = \Delta^N_{q} \frac{\left(\bar{N}\gamma^\mu \gamma_5 N\right) \left( \bar{\chi}\gamma_\mu \gamma_5 \chi\right)}{\Lambda^2}  \,, \nonumber 
\eeq
with~\cite{Belanger:2008sj}
\bea
\Delta_u^p & =& \Delta_d^n \, =\, 0.842\pm 0.012 \,, \nonumber \\
\Delta_d^p & = & \Delta_u^n \,=\, -0.427\pm 0.013 \,, \nonumber \\
\Delta_s^p & = & \Delta_s^n\, = \,-0.085\pm 0.018   \,.
\eea
The total cross section  is then
\beq
\sigma^{Nq}_{3}\,= \,\frac{3\, \mu^2}{\,\pi\, \Lambda^4} \, (\Delta_{q}^{N})^2 \,.
\eeq
The Tevatron limits on spin dependent dark matter scattering for the various operators are shown in Figure~\ref{fig-SD-limits} along with limits from XENON10~\cite{Angle:2007uj}, COUPP~\cite{Behnke:2008zza}, PICASSO~\cite{Archambault:2009sm}
 and ZEPLIN III~\cite{Lebedenko:2009xe}.
For the DM-proton spin-dependent scattering cross section (left panel) we have found that the Tevatron limits are stronger than any other direct detection experiments for all three operators. For the DM-neutron scattering in the right panel, the Tevatron limit is still the best for the up-type quark operator. Limits for a flavor universal operator are close to those of the pure up operator.
%
\subsection{Inelastic, Exothermic and streams}

We will now discuss constraints on several dark matter scenarios which have been proposed in the context of the DAMA modulation signal.  The event rate in a direct detection experiment at a given recoil energy $E_R$ is proportional to 
\begin{equation}
\label{eq-DMrate}
\frac{dR}{dE_R}\propto  n_\chi \sigma_N \int_{v_{min}}^{v_{esc}}\frac{f(v)}{v}dv\,,
\end{equation}
where $f(v)$ is the velocity distribution of dark matter and $n_\chi$ is the number density of the dark matter species in question.

One interesting possibility is that if dark matter up-scatters from a ground state to a slightly excited one, the minimum velocity required for scattering is affected as
\begin{equation}
v_{min} = \sqrt{\frac{1}{2m_T E_R}}\left(\frac{m_T E_R}{\mu_{T}}+\delta\right)
\end{equation}
where $m_T$ is the target nucleus mass, $\mu_T$ is the target-dark matter reduced mass, and $\delta$ is the mass splitting. This increase in the minimum velocity required to scatter causes the integral in equation~(\ref{eq-DMrate}) to be suppressed in iDM models (it also becomes more sensitive to velocity modulation, thus explaining DAMA well).
In order to keep a certain event rate fixed, one would need to enhance the nucleon cross section by the corresponding factor. However, high energy collisions are not sensitive to the $O(100\mbox{ keV})$ splittings between the dark matter states and thus collider bounds will have a relative advantage here. 

\begin{figure}[t]
\begin{center}
\includegraphics[width=0.48\textwidth]{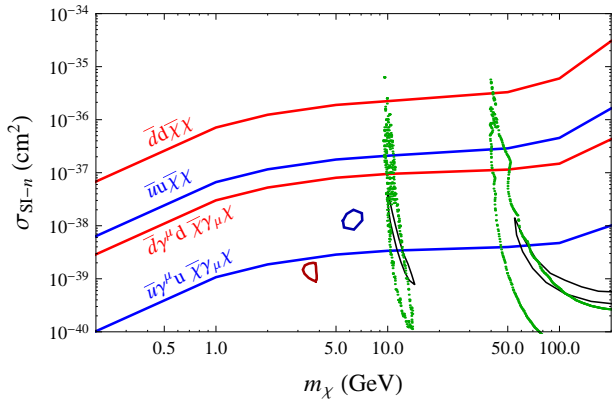}
\caption{The constraints on iDM and exoDM with the DAMA best fit region for iDM as given in Ref.~\cite{Kopp:2009qt} (green, dotted), and in Ref.~\cite{Chang:2008gd, Chang:2008xa} (black, solid).  The DAMA best fit regions for exothermic DM taken from Ref.~\cite{Graham:2010ca} is shown in red (small lower region) and from Ref.~\cite{Essig:2010ye} in blue (small upper region), note that this model uses a light mediator, which will weaken the mono-jet constraints.
}
\label{fig-iDM-limits}
\end{center}
\end{figure}

To explore the sensitivity of the mono-jet search to iDM we show the spin independent bounds again in Figure~\ref{fig-iDM-limits} now compared to the parameter space of iDM that explains DAMA at 99\% C.L., taken from several studies. The DAMA preferred region depends sensitively on various assumptions to which our bounds are not sensitive. These range from astrophysical quantities such as the dark matter velocity distribution as well as experimental issues such as the DAMA energy resolution and the presence or absence of channeling in DAMA. In particular in Figure~\ref{fig-iDM-limits} we show the DAMA best fit region in the black contour region from~Ref.~\cite{Chang:2008gd, Chang:2008xa} at 99\% C.L. for a fixed value $\delta = 35$~keV (for the lower mass region, corresponding to scattering off sodium) and $\delta=120$~keV (for the higher mass region, corresponding to scattering off iodine). We also included another fit in the green dashed region from~Ref.~\cite{Kopp:2009qt}, where the parameter $\delta$ has been treated as a floating parameter.  All of these regions are also constrained by other direct detection experiments to various degrees, depending on the assumptions. We emphasize that the Tevatron is sensitive to a large part of the DAMA preferred region independent of these assumptions. 

Another class of models proposed to explain DAMA is exothermic dark matter~\cite{Graham:2010ca, Essig:2010ye} which involves a light dark matter state which is de-excited upon collision with a nucleus. Here a larger cross section is required in order to enhance the modulation signal which is reduced in exothermic reactions. The allowed region from~\cite{Graham:2010ca} lying below $10^{-39}$ cm$^2$ may also be probed in future Tevatron analyses. 

A third class of models in which collider bounds have a relative advantage are those in which direct detection signals are arising from a sub-dominant component of the dark matter halo. For example, one could exploit the fact that the number density of a thermal relic generically scales as the inverse of the annihilation cross section $n\propto 1/\langle \sigma v \rangle$ to argue that the rate count  at a direct detection experiment, $n\sigma v$ is a constant as the coupling strength of the thermal relic with matter is increased. As the relic couples more strongly it becomes less abundant while keeping the rate fixed.
The Tevatron bound, which obviously does not depend on the number density of dark matter places a direct upper bound on $\sigma$. For example, in~\cite{Essig:2010ye} a light relic which accounts for about 1\% of the dark matter energy density scatters with a cross section of a few$\times10^{-38}$  cm$^2$, which is strongly constrained by the Tevatron mono-jet bound. 

In addition, some direct detection models for DAMA have relied on streams of dark matter that have a low velocity dispersion, but are sub-dominant contribution to the halo density. 
A stream model may also have an enhanced cross section, if the number density in the stream is below that of the generic halo. Examples of such possibilities that may be constrained by our bounds were explored in~\cite{Chang:2008xa,Lang:2010cd}.

It should be emphasized however, that all of these models may escape the Tevatron bound if the interaction of the thermal relic with matter is mediated by a light state, as we shall see below. One may turn this statement around to conclude that if a dark matter model is directly detected in a region that violates the collider bound, then the dark sector is not simply a DM state, but contains a new light mediator through which the DM interacts with the standard model.

In addition to considering the bounds on one operator at a time there are many models that predict more than one operator has non-zero coefficient.  In these cases the direct detection cross section bounds presented above must be combined.  If the direct detection cross section for a single quark flavor, $q$, is $\sigma_q=d_q \mu^2/\Lambda_q^4$ then in the case of multiple operators coupling quarks to DM the direct detection cross section is,
\be
\sigma=\frac{\sum_i d_i}{\sum_i \dfrac{d_i}{\sigma_i}}~.
\ee

\section{Constraints on light mediators} \label{sec:lightmediator}

In placing the bounds in the previous sections we have imagined that the only accessible state from the dark sector is the DM itself, all other states associated with the dark sector are heavy~\cite{Beltran:2010ww}.  However, for certain operators the cutoff scale, shown in Figure~\ref{fig-cutoff}, is low enough to be probed at the Tevatron.  In these situations it may be possible to produce the mediator that generates the four fermion operator directly.  If the mediator couplings with the SM and the dark sector are weak, \ie\ $\le \mathcal{O}(1)$, then the mediator mass is lower than the cutoff scale shown in Figure~\ref{fig-cutoff}, further motivating consideration of mediators within the Tevatron's reach.  Furthermore, recent cosmic ray excesses may be explained by a dark sector that contains a light mediator, $M\sim 1\ \gev$, see for instance \cite{ArkaniHamed:2008qn}.

\begin{figure}[t]
\begin{center}
\includegraphics[width=0.48\textwidth]{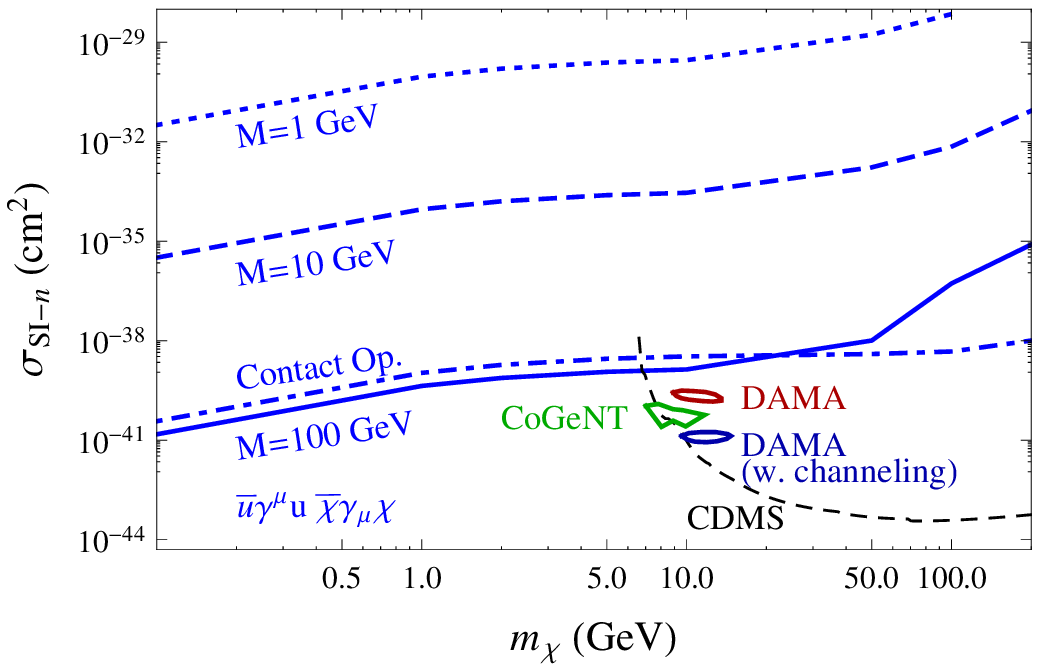}\hspace{2mm}
\includegraphics[width=0.48\textwidth]{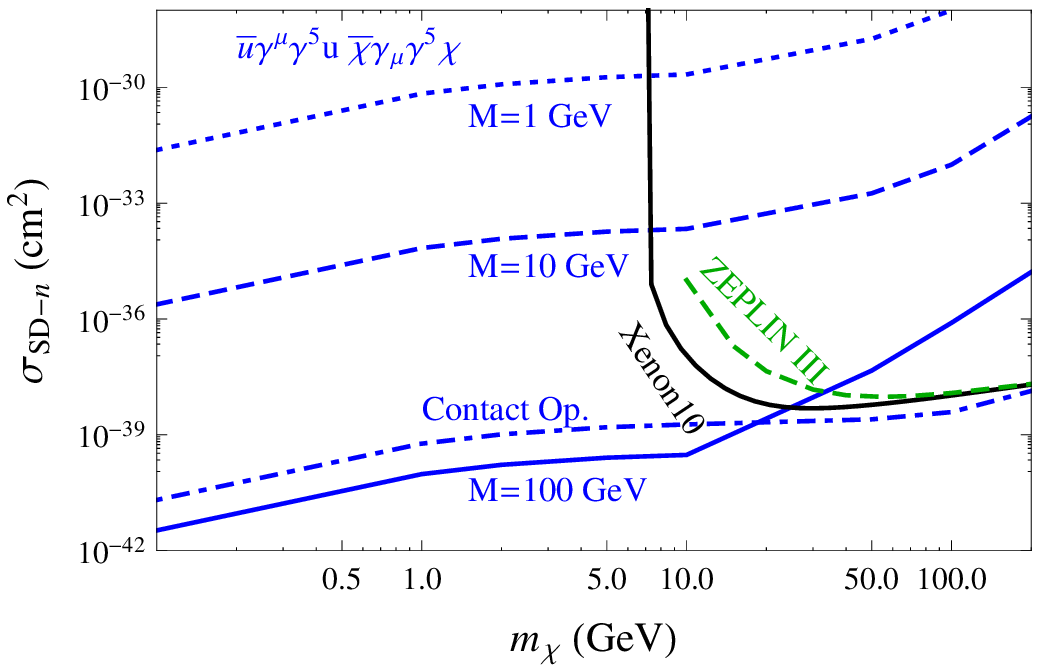}
\caption{Left panel: constraints on the spin-independent DM-neutron scattering cross sections for different mediator masses. Right panel: the same as the left panel but for the constraints on the spin-dependent DM-neutron scattering cross section.}
\label{fig-Mediator-limits}
\end{center}
\end{figure}

As discussed in the introduction, the ratio of the direct detection cross section to the mono-jet production cross section is proportional to $1/M^4$ when the mediator is light.  So, as the mediator mass decreases, the constraints on the dark matter direct-detection from the mono-jet searches become weaker.  For a sufficiently light weakly coupled mediator that satisfies the mono-jet bound the constraint on direct detection will not be competitive with those from direct detection experiments themselves.  However, there is an interesting regime with both a light mediator and light DM, $2m_\chi<M<s^{1/2}$, where the production of DM in mono-jet events can take place through an on-shell mediator which in turn can decay to dark matter.  In this situation the production of dark matter is a two body process rather than three body and so is enhanced by a phase space factor. Note that in this case the mediator could also have a substantial branching fraction to jets, leading to a di-jet invariant mass peak, though this is model dependent and will not be considered further here. 

For simplicity, we consider the mediator to be a SM singlet vector or scalar and consider the effects as its mass is lowered.  In particular, we consider the case of both a $10\ \gev$ and $100\ \gev$ vector mediator, in both cases we consider the width to be 1\% \ie\  $\Gamma=0.01 M$.  We leave the possibility of mediators that are charged under the SM, such as squarks, for future study (though their masses cannot be lowered below current direct bounds). 

As we alter the mass of the mediator we also alter its couplings to the SM and the dark sector, $g_q$ and $g_\chi$ respectively, so that the mono-jet production cross section satisfies the CDF bound.  The effects of a light mediator for the case of SI DM, $\mathcal{O}_2$, and SD DM, $\mathcal{O}_{3}$, are shown in Figure~\ref{fig-Mediator-limits}.  The weakening of the limits for light mediators is clearly seen, and the slight enhancement for the case where the mediator is produced on shell ($M=100\ \gev$ and $m_\chi< 50\ \gev$) is also observable.

\subsection{Momentum dependent}\label{sec:momentumdep}

A particular model of dark matter that requires the introduction of light mediators is the case of DM that has a momentum dependent coupling to nucleons~\cite{Feldstein:2009tr,Chang:2009yt}.
These types of models provide a possible explanation for the DAMA modulation signal, but in order to do so require mediators less than $10\ \gev$ in mass.  Although light from a collider perspective the masses considered are still sufficiently heavy that at direct detection experiments the mediator can be integrated out and an effective four fermion operator can be written.  The axial-scalar operator $\mathcal{O}_4$ leads to momentum dependent and spin dependent dark matter scattering and at the nucleon level the operator is,
\be
\mathcal{O}_4^{Nq} =  -i\, C^N_q \frac{\left(\bar{N}\gamma_5 N\right) \left( \bar{\chi}\gamma_5\chi\right)}{\Lambda^2}~,
\ee
where we have integrated out the mediator and $\Lambda=M/\sqrt{g_q\, q_\chi}$.  In going from quark to nucleonic operators we introduce $C_q^N=\bra{N} \bar{q} i\gamma_5 q \ket{N}$, which are determined in Ref.~\cite{Cheng:1988im} (under the assumption that the large $N_c$ limit is a reasonable approximation of QCD),
\bea
C_u^p &=  141\pm 34,\,            & C_u^n = -138\pm34         \,,       \nonumber \\
C_d^p & = -137\pm33,\,           & C_d^n = 139\pm34         \,,     \nonumber \\
C_s^p & =  -4.15\pm1.0,\,           & C_s^n  = - 0.94\pm0.33  \,.
\label{eq:matrixelementaxialscalar}
\eea

\begin{figure}[t]
\begin{center}
\includegraphics[width=0.5\textwidth]{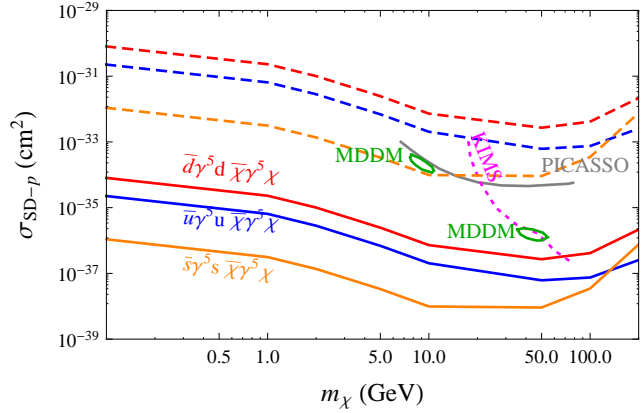}
\caption{The constraints on the momentum and spin dependent model from mono-jet searches. The  solid lines are for a mediator with $M = 10$~GeV, while the dashed lines are a mediator with $M=1$~GeV. The DAMA allowed region is shown in the green contours and is taken from Ref.~\cite{Chang:2009yt}.}
\label{fig-Form-limits}
\end{center}
\end{figure}

The differential cross section for DM scattering off a nucleon is given by
\be
\frac{d\sigma_4^{Nq}}{d\cos\theta} = \frac{1}{32\pi \Lambda^4} \, \frac{q^4}{(m_\chi+m_N)^2}\, (C^N_q)^2 \,,
\ee
where $q$ is the exchanging momentum of the DM scattering off the nucleon. 

Following Ref.~\cite{Chang:2009yt}, we use a reference momentum, $q_{\rm ref} = 100$~MeV, and compare the Tevatron constraints to the region of parameter space that best fits the DAMA result, taken from Figure 3(b) in~\cite{Chang:2009yt}).  The results are shown in Figure~\ref{fig-Form-limits}; we consider the cases of $M=1,\,10\ \gev$.   

We see that the dilution of the Tevatron constraints by the light mediator means that momentum dependent dark matter with $M=1\ \gev$ is not severely constrained by the mono-jet search.  However, if instead the mediator is 10 GeV and has $\mathcal{O}(1)$ couplings, then the lack of a mono-jet excess places strong constraints on the model and rules out the DAMA preferred region\footnote{This option may well be ruled out by other limits.}, note that unlike previous cases, the constraints coming from the strange quarks are the most stringent. This is due to a small matrix element for the strange quark in equation~(\ref{eq:matrixelementaxialscalar}).

\section{Discussions and conclusions}\label{sec-discussion}

\begin{figure}[t]
\begin{center}
\includegraphics[width=0.48\textwidth]{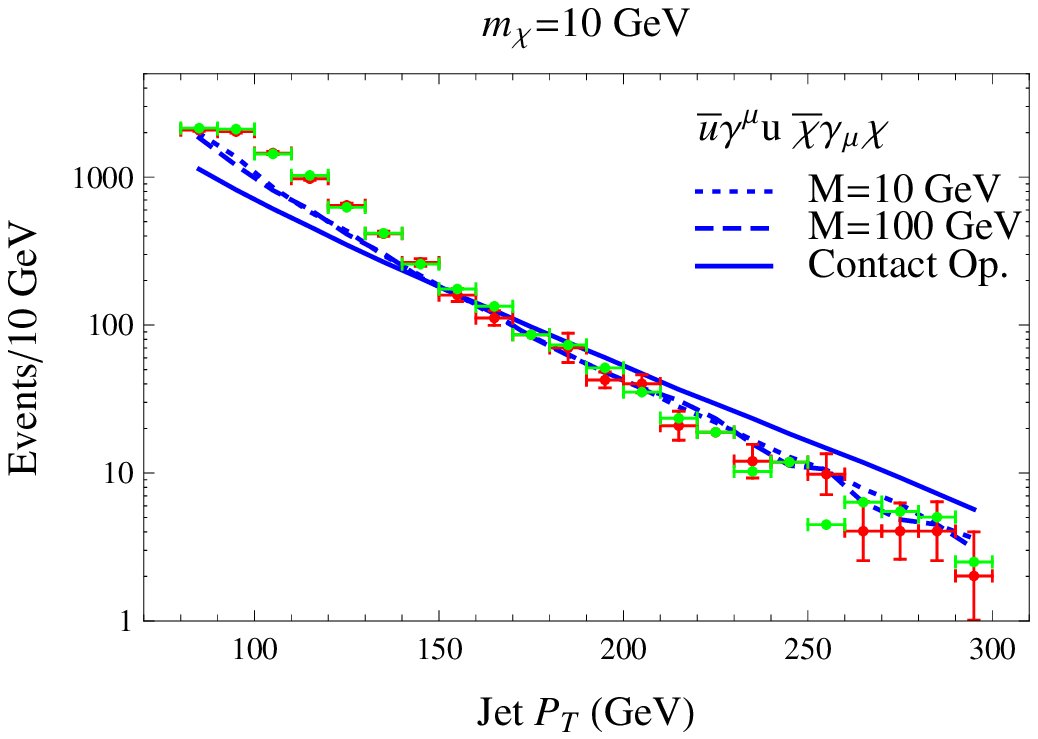} \vspace{2mm}
\includegraphics[width=0.48\textwidth]{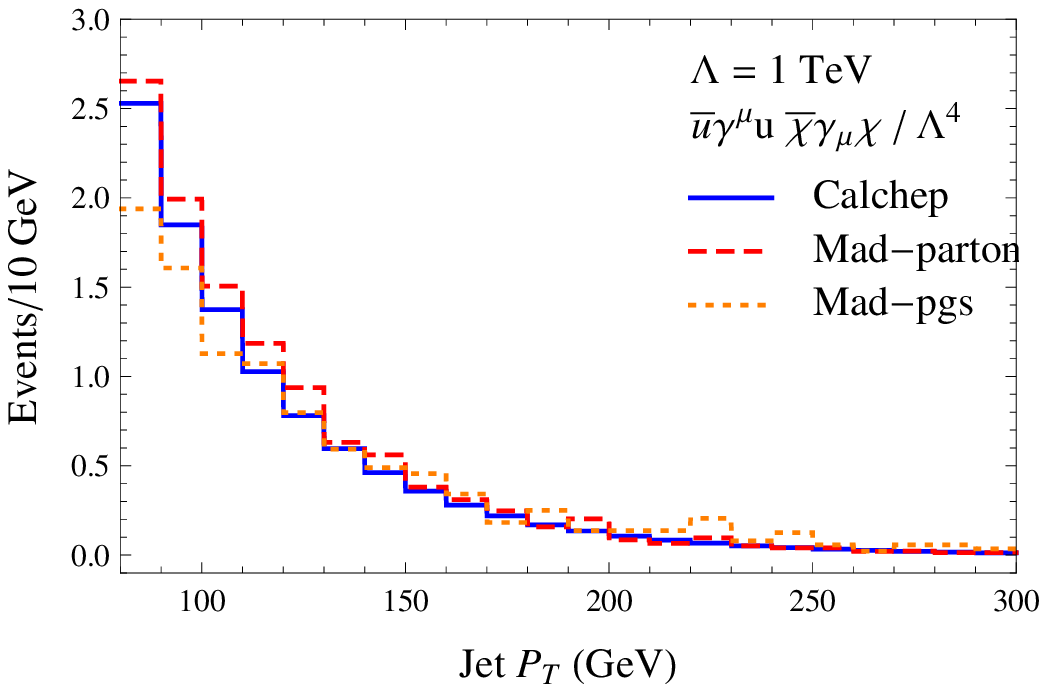}\\
(a)\qquad\qquad\qquad\qquad\qquad\qquad\qquad\qquad\qquad\qquad(b)
\caption{(a) Comparisons of the shapes of the signal, the SM background and CDF measured events. The SM predictions are shown in the green and the CDF observed data are shown in red. (b) Comparisons of simulated signal events from two different Monte-Carlo tools and for the parton and the particle levels. The cutoff $\Lambda \equiv M/\sqrt{g_\chi g_q}$ is chosen to be 1 TeV.}
\label{fig-spectra}
\end{center}
\end{figure}

It is worthwhile to consider possible improvements to the dark matter search at the Tevatron, and in the future at the LHC. Here we placed bounds on dark matter using only the total rate of mono-jet signal events above a certain $p_T$ cut. An analysis that takes the spectrum shape into account may yield more powerful bounds.  
We show the spectral shape of the signal compared to the background in Figure~\ref{fig-spectra}(a). We find that the signal spectrum is somewhat harder than the background, especially when the messenger mass is much higher than the dark matter mass. 
We find that including showering, hadronization (using Pythia~\cite{Sjostrand:2006za}) and a detector simulation (PGS~\cite{PGS}) does not change the signal shape significantly, particularly above 100 GeV, as is shown in Figure~\ref{fig-spectra}(b).
This may allow us to place tighter constraints using a multi-bin analysis as compared with a simple counting experiment, since signal predicts more deviations in high $p_T$ bins. However, this would require knowledge of the theoretical uncertainty on a bin-by-bin basis which is not presently available.  Furthermore, a bound may be extracted from mono-photon events.

In this work we show that the Tevatron mono-jet search places competitive bounds on dark matter-nucleus cross sections relevant for direct detection experiments. 
In particular, the Tevatron limits are the current world-best for light dark matter, below a mass of 5 GeV. The Tevatron also sets the best limit spin dependent dark matter scattering. Various models built to explain the DAMA modulation signal such as inelastic and exothermic dark matter are also constrained by current Tevatron searches.  

In addition to considering dark matter that couples to quarks via contact interactions we have taken the possibility of light mediators, as motivated by cosmic ray excesses~\cite{ArkaniHamed:2008qn} into account. We find that the introduction of a light mediator of mass $\lesssim 10$ GeV alleviates the mono-jet bounds completely for most cases.  This leads to an interesting conclusion - if a direct dark matter signal is established in a region that is in conflict with collider bounds, a new light state should be introduced to reconcile the data.  The Tevatron is unable to search for DM above a few 100 GeV due to kinematics, an upper bound that will be raised at the LHC. The current powerful bounds using only 1 fb$^{-1}$ motivates a dedicated analysis using more Tevatron data as well as future analyses at the LHC.

\vspace{0.5cm}
\textbf{Note added:} While in the final stages of this work Ref.~\cite{Goodman:2010yf} appeared on the arXiv.  They address similar issues, focussing on Majorana DM, but do not consider light mediators.

\subsection*{Acknowledgements} 
We would like to thank John Campbell, Walter Giele, Joachim Kopp, Adam Martin and Neal Weiner for useful discussions.  We are grateful to Joachim Kopp for providing the contours used in Figure~\ref{fig-iDM-limits}.  PJF would like to thank the CCCP at NYU, and the UC Davis HEFTI workshop on Light Dark Matter for hospitality while part of this work was undertaken.
Fermilab is operated by Fermi Research Alliance, LLC, under Contract DE-AC02-07CH11359 with the United States Department of Energy. 

\appendix
\section{Matrix Elements}

We present here analytic expressions for the mono-jet processes mediated by the operators in equation (\ref{eq:Ops}).  We consider only the case of the mediator being sufficiently massive it may be integrated out and we take the initial state quarks as massless and label momenta as $q(p_1)+\bar{q}(p_2)\rightarrow g(q)+ \chi(k_1) + \chi(k_2)$.  In the center of mass frame, the differential cross section for mono-jet production is
\be
\frac{d\sigma_i}{dE_1 dE_2 d\cos\theta d\phi d\gamma}=\frac{1}{512\pi^5 E_{cm}^2}|\mathcal{M}_i|^2~,
\ee
where we have introduced final state energies and angles through,
\bea
k_1 &=& \sqrt{E_1^2-m_\chi^2}\left(\frac{E_1}{\sqrt{E_1^2-m_\chi^2}},\sin\theta\cos\phi,\sin\theta\sin\phi,\cos\theta\right)~, \nonumber \\
k_2 &=&  \sqrt{E_2^2-m_\chi^2}\left(\frac{E_2}{\sqrt{E_2^2-m_\chi^2}},\sin\theta\cos(\phi+\alpha),\sin\theta\sin(\phi+\alpha),\cos\theta\right)~.
\eea
The matrix element for each operator is,
\bea
|\mathcal{M}_1|^2 & = & \frac{16}{9 \Lambda^4} \frac{\left(k_1.k_2-m_\chi^2\right) p_2.q}{p_1.q}~, \\
|\mathcal{M}_2|^2 & = & \frac{32}{9 \Lambda^4}\frac{ (k_2.p_2)(k_1. q) + (k_1. p_2)(k_2.q) +  m_\chi^2 (p_2.q) }{ p_1.q}~, \\
|\mathcal{M}_3|^2 & = & \frac{32}{9 \Lambda^4}\frac{(k_2.p_2)(k_1.q) + (k_1.p_2)(k_2.q) -  m_\chi^2 (p_2.q)}{p_1.q}~, \\
|\mathcal{M}_4|^2 & = & \frac{16}{9 \Lambda^4}\frac{\left(k_1.k_2 + m_\chi^2\right) p_2.q}{p_1.q} ~.
\eea
As before we define $\Lambda=M/\sqrt{g_\chi g_q}$.
\section{Relic abundance}

\begin{figure}[t]
\begin{center}
\includegraphics[width=0.48\textwidth]{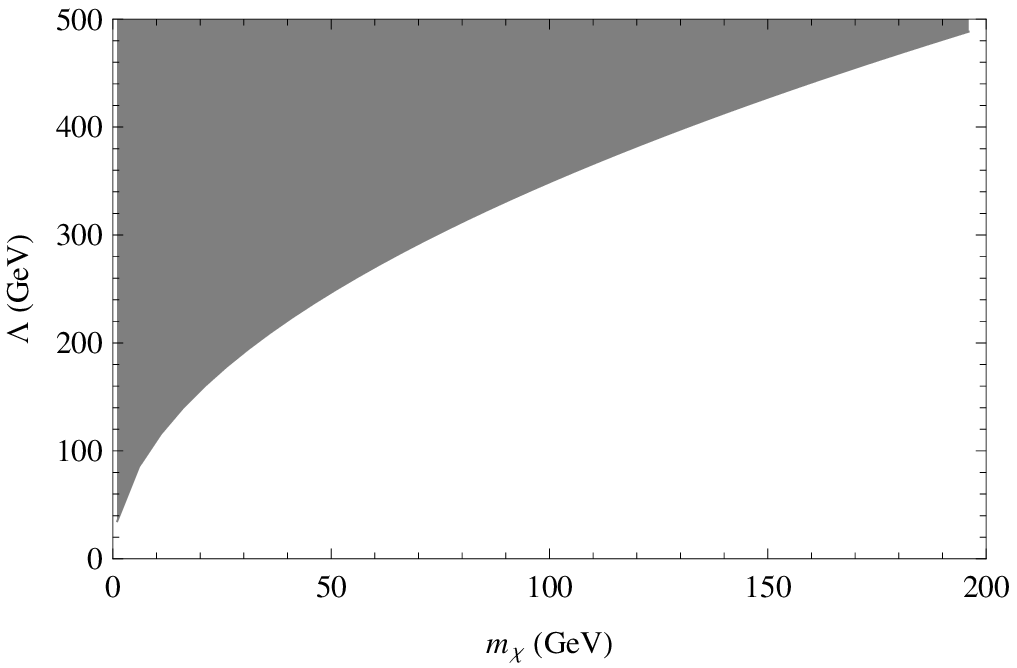} \hspace{2mm}
\includegraphics[width=0.48\textwidth]{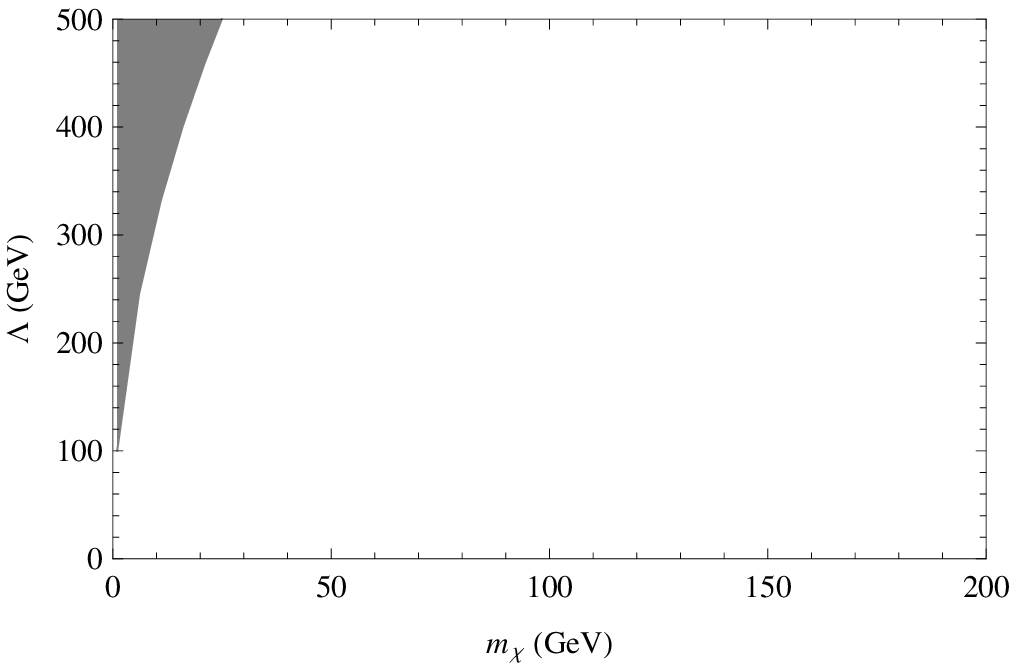} \\ \vspace{2mm}
\includegraphics[width=0.48\textwidth]{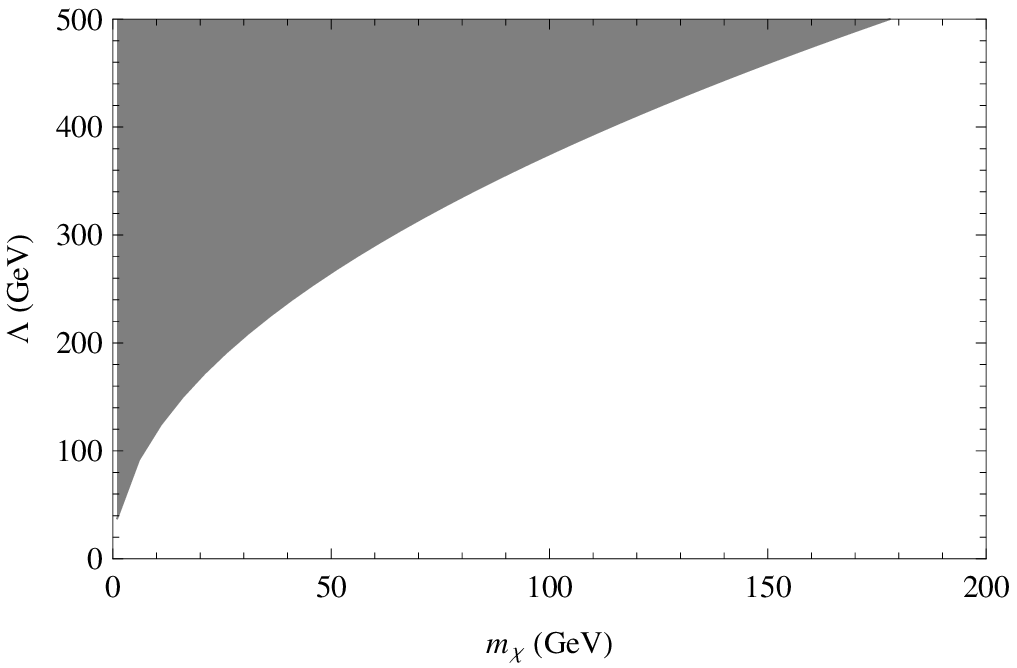} \hspace{2mm}
\includegraphics[width=0.48\textwidth]{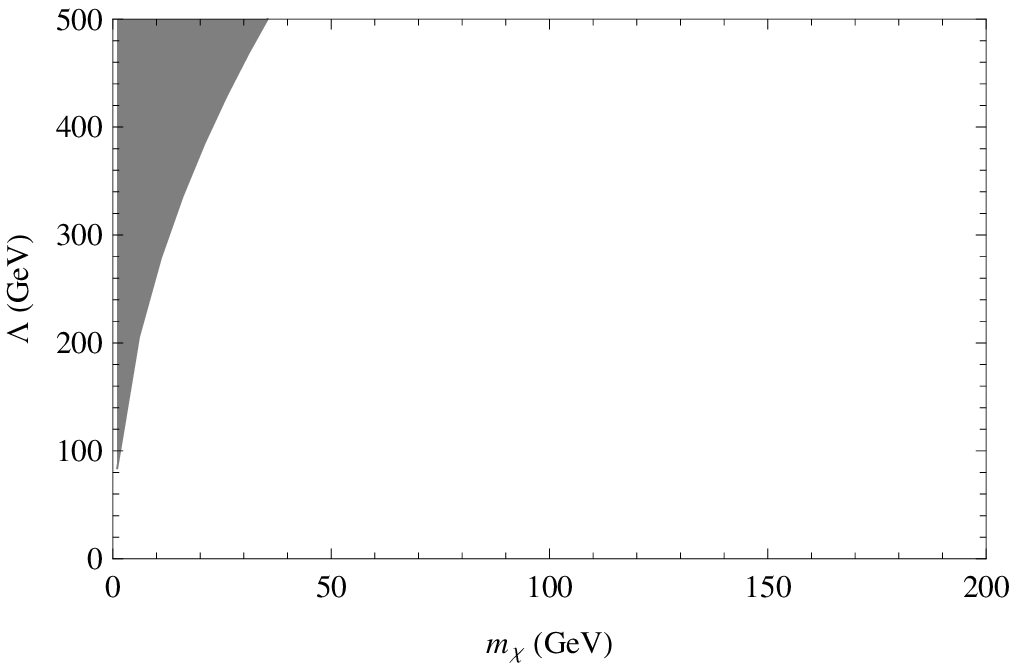}\\ \hspace{2mm}
\caption{Abundance of DM if it were a thermal relic, the shaded region has abundance greater than the WMAP bound, $\Omega_\chi h^2 = 0.1120$.  Reading from top left to bottom right the plots are for $\mathcal{O}_1, \mathcal{O}_2, \mathcal{O}_3 ,\mathcal{O}_4$, and in each case we have considered only the coupling to one flavor of light quarks.}
\label{fig-abundance}
\end{center}
\end{figure}
The DM abundance as measured by WMAP is $\Omega_\chi h^2 = 0.1120 \pm 0.0056$.  If the DM is a thermal relic its abundance after freeze-out can be related to its annihilation cross section~\cite{Hooper:2009zm} through,
\be
\Omega h^2 \approx \frac{1.04\times 10^{9}\ \gev x_F}{M_{pl}\sqrt{g_*}(a+3b/x_F)}~,
\ee
where the partial wave expansion of the annihilation cross section is $\sigma v = a + b v^2 + \ldots$.  For a typical weak scale WIMP $g_*\approx 100$ and $x_F=m_\chi/T_F \approx 20-30$ and the annihilation cross section should be $\sigma v \sim 3\times 10^{-26}\mathrm{cm}^3\mathrm{s}^{-1}$.  For each of the operators under discussion the annihilation cross section, expanded in relative velocity, $v$, is
\bea
\sigma_1 v & = & \frac{3}{8\pi\Lambda^4}\sqrt{1-\frac{m_q^2}{m_\chi^2}}(m_\chi^2-m_q^2)\, v^2~, \\
\sigma_2 v & = & \frac{1}{16\pi\Lambda^4}\sqrt{1-\frac{m_q^2}{m_\chi^2}} \left(24 (2m_\chi^2+m_q^2) + \frac{8 m_\chi^4-4m_\chi^2 m_q^2+5m_q^4}{m_\chi^2-m_q^2}\,v^2\right)~, \\
\sigma_3 v & = & \frac{1}{16\pi\Lambda^4}\sqrt{1-\frac{m_q^2}{m_\chi^2}}\left( 24  m_q^2  +  \frac{8 m_\chi^4-22m_\chi^2 m_q^2+17m_q^4}{(m_\chi^2-m_q^2)}\,v^2\right)~, \\
\sigma_4 v & = & \frac{3}{16\pi\Lambda^4}\sqrt{1-\frac{m_q^2}{m_\chi^2}}m_\chi^2\left(8 + \frac{2 m_\chi^2- m_q^2}{m_\chi^2-m_q^2}\,v^2\right)~.
\eea
We show the resulting relic abundance if the DM were a thermal relic in Fig.~\ref{fig-abundance}.  It should be noted that the DM need not be a thermal relic, this is a model dependent assumption, and our model independent approach is sensitive to all regions.  In particular, regions where the thermal relic abundance is below the WMAP bound are still viable regions to be probed by direct detection and collider searches.  Furthermore, in the region where the abundance is above that of WMAP there may be additional annihilation modes, not useful for searches at the Tevatron, that would lower the abundance.

\providecommand{\href}[2]{#2}\begingroup\raggedright\endgroup

\end{document}